\documentclass{article}
\usepackage{graphicx,amssymb}

\usepackage{amsmath}

\newtheorem{theorem}{Theorem}

\newtheorem{definition}[theorem]{Definition}

\newtheorem{fprop}[theorem]{Proposition}
\begin{document}
%\begin{frontmatter}

\title{New quantum algorithm for studying NP-complete problems}
\author{Masanori Ohya
\\
Tokyo University of Science,\\
Department of Information Sciences,\\
Noda City, Chiba 278-8510, Japan \\
{\small e-mail: ohya@is.noda.tus.ac.jp}
\medskip
\\
%\corauth[cor]{Corresponding author.}
%\ead{ohya@is.noda.tus.ac.jp}
Igor V. Volovich \\
Steklov Mathematical Institute,\\
Gubkin St. 8, 117966 Moscow,
Russia \\
{\small e-mail: volovich@mi.ras.ru}}

%EndAName
%\ead{volovich@mi.ras.ru}
%\address%[MO]
%{ ^*
%Tokyo University of Science,
%Department of Information Sciences,
%Noda City, Chiba 278-8510, Japan \\
%e-mail: ohya@is.noda.tus.ac.jp}
%\address%[IV]
%{ ^**
%Steklov Mathematical Institute,
%Gubkin St. 8, 117966 Moscow,
%Russia \\
%e-mail: volovich@mi.ras.ru}
%email:ohya@is.noda.tus.ac.jp\\
%volovich@mi.ras.ru}
\maketitle

\begin{abstract}
Ordinary approach to quantum algorithm is based on quantum Turing machine or
quantum circuits. It is known that this approach \ is not powerful enough to
solve NP-complete problems. In this paper we study a new approach to quantum
algorithm which is a combination of the ordinary quantum algorithm with a
chaotic dynamical system. \ We consider the satisfiability problem as an
example of NP-complete problems and argue that the problem, in principle,
can be solved in polynomial time by using our new quantum algorithm.\\
{\bf keywords:}
Quantum Algorithm, NP-complete problem, Chaotic Dynamics
\end{abstract}
%\begin{keyword}
%Quantum Algorithm, NP-complete problem, Chaotic Dynamics
%\PACS 03.67.Lx, 05.45.Ac
%\end{keyword}
%\end{frontmatter} 

\section{Introduction}

Ordinary approach to quantum algorithm is based on quantum Turing machine or
quantum circuits \cite{BEZ,Deu,BV}. It is known that this approach \ is not
powerful enough to solve NP-complete problems \cite{GJ,BBBV}. In \cite{OV1}
we have proposed a new approach to quantum algorithm which goes beyond the
standard quantum computation paradigm. This new approach is a sort of
combination of the ordinary quantum algorithm and a chaotic dynamics. This
approach was based on the results obtained in the paper \cite{OM}.

There are important problems such as the knapsack problem, the traveling
salesman problem, the integer programming problem, the subgraph isomorphism
problem, the satisfiability problem that have been studied for decades and
for which all known algorithms have a running time that is exponential in
the length of the input. These five problems and many other problems belong
to the set of \textbf{NP}-complete problems \cite{GJ}.

Many \textbf{NP}-complete problems have been identified, and it seems that
such problems are very difficult and probably exponential. If so, solutions
are still needed, and in this paper we consider an approach to these
problems based on quantum computers and chaotic dynamics as mentioned above.

As in the previous papers \cite{OM,OV1}\ we again consider the
satisfiability problem as an example of NP-complete problems and argue that
the problem, in principle, can be solved in polynomial time by using our new
quantum algorithm.

It is widely believed that quantum computers are more efficient than
classical computers. In particular Shor \cite{Sho,EJ} gave a remarkable
quantum polynomial-time algorithm for the factoring problem. However, it is
known that this problem is not \textbf{NP}-complete but is NP-intermidiate.

Since the quantum algorithm of the satisfiability problem (SAT for short)
has been considered in \cite{OM}, Accardi and Sabbadini showed that this
algorithm is combinatric one and they discussed its combinatric
representation \cite{AS}.\ It was shown in \cite{OM} that the SAT problem
can be solved in polynomial time by using a quantum computer under the
assumption that a special superposition of two orthogonal vectors can be
physically detected . The problem one has to overcome here is that the
output of computations could be a very small number and one needs to amplify
it to a reasonable large quantity.

In this paper we construct a new model (representation) of computations
which combine ordinary quantum algorithm with a chaotic dynamical system and
prove that one can solve the SAT problem in polynomial time.

For a recent discussion of computational complexity in quantum computing see 
\cite{FR,Cle,HHZ,AL}. Mathematical features of quantum computing and quantum
information theory are summarized in \cite{OV3}.

\section{SAT Problem}

Let $X\equiv \left\{ x_{1},\cdots ,x_{n}\right\} $ be a set. Then $x_{k}$
and its negation $\overline{x}_{k}$ $(k=1,2,\cdots ,n)$ are called literals
and the set of all such literals is denoted by $X^{^{\prime }}=\left\{ x_{1},%
\overline{x}_{1},\cdots ,x_{n},\overline{x}_{n}\right\} $. The set of all
subsets of $X^{^{\prime }}$ is denoted by $\mathcal{F}(X^{^{\prime }})$ and
an element $C\in \mathcal{F}(X^{^{\prime }})$ is called a clause. We take a
truth assignment to all variables $x_{k}.$ If we can assign the truth value
to at least one element of $C,$ then $C$ is called satisfiable. When $C$ is
satisfiable, the truth value $t(C)$ of $C$ is regarded as true, otherwise,
that of $C$ is false. Take the truth values as true $"1",$ false$\ "0".$
Then 
\begin{equation*}
C\text{ is satisfiable iff }t(C)=1.
\end{equation*}

Let $L=\left\{ 0,1\right\} $ be a Boolean lattice with usual join $\vee $
and meet $\wedge ,$ and $t(x)$ be the truth value of a literal $x$ in $X.$
Then the truth value of a clause $C$ is written as

\begin{equation*}
t(C)\equiv \vee _{x\in C}t(x).
\end{equation*}
Further the set $\mathcal{C}$ of all clauses $C_{j}$ $(j=1,2,\cdots m)$ is
called satisfiable iff the meet of all truth values of $C_{j}$ is 1;

\begin{equation*}
t(\mathcal{C})\equiv \wedge _{j=1}^{m}t(C_{j})=1.
\end{equation*}
\ 

Thus the SAT problem is written as follows:

\begin{definition}
SAT Problem: Given a set $X\equiv \left\{ x_{1},\cdots ,x_{n}\right\} $ and
a set $\mathcal{C=}\left\{ C_{1},C_{2},\cdots ,C_{m}\right\} $ of clauses,
determine whether $\mathcal{C}$ is satisfiable or not.
\end{definition}

That is, this problem is to ask whether there exsits a truth assignment to
make $\mathcal{C}$ satisfiable.

It is known\cite{GJ} in usual algorithm that it is polynomial time to check
the satisfiability only when a specific truth assignment is given, but we
can not determine the satisfiability in polynomial time when an assignment
is not specified.

Note that a formula made by the product (AND $\wedge $) of the disjunction
(OR $\vee $) of literals is said to be in the \textit{product of sums }(POS)
form. For example, the formula 
\begin{equation*}
\left( x_{1}\vee \overline{x}_{2}\right) \wedge \left( \overline{x}%
_{1}\right) \wedge \left( x_{2}\vee \overline{x}_{3}\right) 
\end{equation*}
is in POS form. Thus a formula in POS form is said to be \textit{satisfiable 
}if there is an assignment of values to variables so that the formula has
value 1. Therefore the SAT problem can be regarded as determining \emph{%
whether or not a formula in POS form is satisfiable}.

The following analytical formulation of SAT problem is useful. We define a
family of Boolean polynomials $f_{\mathcal{A}}$, indexed by the following
data. One $\mathcal{A}$ is a set 
\begin{equation*}
\mathcal{A}=\left\{ S_{1},...,S_{N},T_{1},...,T_{N}\right\} ,
\end{equation*}
where $S_{i},T_{i}\subseteq \left\{ 1,...,n\right\} ,$ and $f_{\mathcal{A}}$
is defined as 
\begin{equation*}
f_{\mathcal{A}}(x_{1},\cdots ,x_{n})=\prod_{i=1}^{N}\left( 1+\prod_{a\in
S_{i}}(1-x_{a})\prod_{b\in T_{i}}x_{b}\right) .
\end{equation*}
We assume here the addition modulo 2. The SAT problem now is to determine
whether or not there exists a value of $\mathbf{x}=(x_{1},\cdots ,x_{n})$
such that $f_{\mathcal{A}}(\mathbf{x})=1.$

\section{Quantum Algorithm}

Although the quantum algorithm of SAT problem is needed to add the dust bits
to the input $n$ bits, the number of dust bites has been shown the order of $%
n$ \cite{OM,AS}. Therefore for simplicity we will work in the $\left(
n+1\right) $-tuple tensor product Hilbert space $\mathcal{H\equiv }$ $%
\otimes _{1}^{n+1}$\textbf{C}$^{2}$ in this paper with the computational
basis 
\begin{equation*}
\left| x_{1},...,x_{n},y\right\rangle =\otimes _{i=1}^{n}\left|
x_{i}\right\rangle \otimes \left| y\right\rangle
\end{equation*}
where $x_{1},...,x_{n},$ $y=0$ or $1.$ We denote $\left|
x_{1},...,x_{n},y\right\rangle =\left| \mathbf{x},y\right\rangle .$ The
quantum version of the function $f(\mathbf{x}):=f_{\mathcal{A}}(\mathbf{x})$
is given by the unitary operator $U_{f}\left| \mathbf{x},y\right\rangle
=\left| \mathbf{x},y+f(\mathbf{x})\right\rangle .$ We assume that the
unitary matrix $U_{f}$ can be build in the polynomial time, see \cite{OM}.
Now let us use the usual quantum algorithm:

(i) By using the Fourier transform produce from $\left| \mathbf{0,}%
0\right\rangle $ the superposition 
\begin{equation*}
\left| v\right\rangle :=\frac{1}{\sqrt{2^{n}}}\sum_{\mathbf{x}}\left| 
\mathbf{x},0\right\rangle .
\end{equation*}

(ii) Use the unitary matrix $U_{f}$ to calculate $f(\mathbf{x}):$%
\begin{equation*}
\left| v_{f}\right\rangle =U_{f}\left| v\right\rangle =\frac{1}{\sqrt{2^{n}}}%
\sum_{\mathbf{x}}\left| \mathbf{x},f(\mathbf{x})\right\rangle
\end{equation*}
Now if we measure the last qubit, i.e., apply the projector $P=I\otimes
\left| 1\right\rangle \left\langle 1\right| $ to the state $\left|
v_{f}\right\rangle ,$ then we obtain that the probability to find the result 
$f(\mathbf{x})=1$ is $\left\| P\left| v_{f}\right\rangle \right\|
^{2}=r/2^{n}$ where $r$ is the number of roots of the equation $f(\mathbf{x}%
)=1.$ If $r$ is suitably large to detect it, then the SAT problem is solved
in polynominal time. However, for small $r,$ the probability is very small
and this means we in fact don't get an information about the existence of
the solution of the equation $f(\mathbf{x})=1,$ so that in such a case we
need further deliberation.

Let us simplify our notations. After the step (ii) the quantum computer will
be in the state 
\begin{equation*}
\left| v_{f}\right\rangle =\sqrt{1-q^{2}}\left| \varphi _{0}\right\rangle
\otimes \left| 0\right\rangle +q\left| \varphi _{1}\right\rangle \otimes
\left| 1\right\rangle
\end{equation*}
where $\left| \varphi _{1}\right\rangle $ and $\left| \varphi
_{0}\right\rangle $ are normalized $n$ qubit states and $q=\sqrt{r/2^{n}}.$
Effectively our problem is reduced to the following $1$ qubit problem. We
have the state 
\begin{equation*}
\left| \psi \right\rangle =\sqrt{1-q^{2}}\left| 0\right\rangle +q\left|
1\right\rangle
\end{equation*}
and we want to distinguish between the cases $q=0$ and $q>0$(small positive
number).

It is argued in \cite{BBBV} that quantum computer can speed up \textbf{NP}
problems quadratically but not exponentially. The no-go theorem states that
if the inner product of two quantum states is close to 1, then the
probability that a measurement distinguishes which one of the two it is
exponentially small. And one could claim that amplification of this
distinguishability is not possible.

At this point we emphasize that we do not propose to make a measurement (not
read) which will be overwhelmingly likely to fail. What we do it is a
proposal to use the output $\left| \psi \right\rangle $ of the quantum
computer as an input for another device which uses chaotic dynamics in the
sequel.

The amplification would be not possible if we use the standard model of
quantum computations with a unitary evolution. However the idea of our paper
is different. We propose to combine quantum computer with a chaotic dynamics
amplifier. Such a quantum chaos computer is a new model of computations
going beyond usual scheme of quantum computation and we demonstrate that the
amplification is possible in the polynomial time.

One could object that we don`t suggest a practical realization of the new
model of computations. But at the moment nobody knows of how to make a
practically useful implementation of the standard model of quantum computing
ever. Quantum circuit or quantum Turing machine is a mathematical model
though convincing one. It seems to us that the quantum chaos computer
considered in this paper deserves an investigation and has a potential to be
realizable.

In this paper we propose a mathematical model of computations for solving
SAT problem by refining our previous paper \cite{OV1}. A possible specific
physical implementation of quantum chaos computations with some error
correction will be discussed in a separate paper \cite{OV2}, which is some
how related to the recently proposed atomic quantum computer \cite{Vol1}.

\section{Chaotic Dynamics}

Various aspects of classical and quantum chaos have been the subject of
numerious studies, see \cite{O2} and ref's therein.The investigation of
quantum chaos by using quantum computers has been proposed in \cite
{GCZ,Sch,KM}. Here we will argue that chaos can play a constructive role in
computations.

Chaotic behaviour in a classical system usually is considered as an
exponential sensitivity to initial conditions. It is this sensitivity we
would like to use to distinquish between the cases $q=0$ and $q>0$ from the
previous section.

Consider the so called logistic map which is given by the equation 
\begin{equation*}
x_{n+1}=ax_{n}(1-x_{n})\equiv g(x),~~~x_{n}\in \left[ 0,1\right] .
\end{equation*}

\noindent \noindent \noindent The properties of the map depend on the
parameter $a.$ If we take, for example, $a=3.71,$ then the Lyapunov exponent
is positive, the trajectory is very sensitive to the initial value and one
has the chaotic behaviour \cite{O2}. It is important to notice that if the
initial value $x_{0}=0,$ then $x_{n}=0$ for all $n.$

It is known \cite{Deu} that any classical algorithm can be implemented on
quantum computer. Our quantum chaos computer will be consisting from two
blocks. One block is the ordinary quantum computer performing computations
with the output $\left| \psi \right\rangle =\sqrt{1-q^{2}}\left|
0\right\rangle +q\left| 1\right\rangle $. The second block is a computer
performing computations of the \textit{classical} logistic map. This two
blocks should be connected in such a way that the state $\left| \psi
\right\rangle $ first be transformed into the density matrix of the form 
\begin{equation*}
\rho =q^{2}P_{1}+\left( 1-q^{2}\right) P_{0}
\end{equation*}
where $P_{1}$ and $P_{0}$ are projectors to the state vectors $\left|
1\right\rangle $ and $\left| 0\right\rangle .$ This connection is in fact
nontrivial and actually it should be considered as the third block. One has
to notice that $P_{1}$ and $P_{0}$ generate an Abelian algebra which can be
considered as a classical system. In the second block the density matrix $%
\rho $ above is interpreted as the initial data $\rho _{0}$, and we apply
the logistic map as 
\begin{equation*}
\rho _{m}=\frac{(I+g^{m}(\rho _{0})\sigma _{3})}{2}
\end{equation*}
where $I$ is the identity matrix and $\sigma _{3}$ is the z-component of
Pauli matrix on $\mathbb{C}^{2}.$ This expression is different from that of
our first paper \cite{OV1}. To find a proper value $m$ we finally measure
the value of $\sigma _{3}$ in the state $\rho _{m}$ such that

\begin{equation*}
M_{m}\equiv tr\rho _{m}\sigma _{3}.
\end{equation*}
After simple computation we obtain

\begin{equation*}
\rho _{m}=\frac{(I+g^{m}(q^{2})\sigma _{3})}{2},\text{ and }%
M_{m}=g^{m}(q^{2}).
\end{equation*}
Thus the question is whether we can find such a $m$ in polynomial steps of $%
n\ $satisfying the inequality $M_{m}\geqq \frac{1}{2}$ for very small but
non-zero $q^{2}.$ Here we have to remark that if one has $q=0$ then $\rho
_{0}=P_{0}$ and we obtain $M_{m}=0$ for all $m.$ If $q\neq 0,$ the
stochastic dynamics leads to the amplification of the small magnitude $q$ in
such a way that it can be detected as is explained below. The transition
from $\rho _{0}$ to $\rho _{m}$ is nonlinear and can be considered as a
classical evolution because our algebra generated by $P_{0}$ and $P_{1}$ is
abelian.The amplification can be done within atmost 2n steps due to the
following propositions. Since $g^{m}(q^{2})$ is $x_{m}$ of the logistic map $%
x_{m+1}=g(x_{m})$ with $x_{0}=q^{2},$ we use the notation $x_{m}$ in the
logistic map for simplicity.

\begin{fprop}
For the logistic map $x_{n+1}=ax_{n}\left( 1-x_{n}\right) $ with $a$ $\in %
\left[ 0,4\right] $ and $x_{0}\in \left[ 0,1\right] ,$ let $x_{0}\ $be $%
\frac{1}{2^{n}}$ and a set $J\ $be $\left\{ 0,1,2,\cdots ,n,\cdots
2n\right\} .$ If $a$ is $3.71,$ then there exists an integer $m$ in $J$
satisfying $x_{m}>\frac{1}{2}.$
\end{fprop}

Proof: Suppose that there does not exist such $m$ in $J.$ Then $x_{m}\leq 
\frac{1}{2}$ for any $m\in J.$ The inequality $x_{m}\leq \frac{1}{2}$ implies

\begin{equation*}
x_{m}=3.71(1-x_{m-1})x_{m-1}\geq \frac{3.71}{2}x_{m-1}.
\end{equation*}
Thus we have

\begin{equation*}
\frac{1}{2}\geq x_{m}\geq \frac{3.71}{2}x_{m-1}\geq \cdots \geq \left( \frac{
3.71}{2}\right) ^{m}x_{0}=\left( \frac{3.71}{2}\right) ^{m}\frac{1}{2^{n}},
\end{equation*}
from which we get

\begin{equation*}
2^{n+m-1}\geq \left( 3.71\right) ^{m}.
\end{equation*}
According to the above inequality, we obtain

\begin{equation*}
m\leq \frac{n-1}{\log _{2}3.71-1}.
\end{equation*}
Since $\log _{2}3.71\doteqdot 1.8912,$ we have

\begin{equation*}
m\leq \frac{n-1}{\log _{2}3.71-1}<\frac{5}{4}\left( n-1\right) ,
\end{equation*}
which is definitely less than $2n-1$ and it is contradictory to the
statement ''$x_{m}\leq \frac{1}{2}$ for any $m\in J".$ Thus there exists $m$
in $J$ satisfying $x_{m}>\frac{1}{2}.\blacksquare $

\begin{fprop}
Let $a$ and $n$ be the same in the above proposition. If there exists $m_{0}$
in $J$ such that $x_{m_{0}}>\frac{1}{2}$ $,$ then $m_{0}>\frac{n-1}{\log
_{2}3.71}.$
\end{fprop}

Proof: Since $0\leq $ $x_{m}\leq 1,$ we have

\begin{equation*}
x_{m}=3.71(1-x_{m-1})x_{m-1}\leq 3.71x_{m-1},
\end{equation*}
which reduces to 
\begin{equation*}
x_{m}\leq \left( 3.71\right) ^{m}x_{0}.
\end{equation*}
For $m_{0}$ in $J$ satisfying $x_{m_{0}}>\frac{1}{2}$ , it holds

\begin{equation*}
x_{0}\geq \frac{1}{\left( 3.71\right) ^{m_{0}}}x_{m_{0}}>\frac{1}{2\times
\left( 3.71\right) ^{m_{0}}}.
\end{equation*}
It follows that from $x_{0}=\frac{1}{2^{n}}$

\begin{equation*}
\log _{2}2\times \left( 3.71\right) ^{m_{0}}>n,
\end{equation*}
which implies

\begin{equation*}
m_{0}>\frac{n-1}{\log _{2}3.71}.\blacksquare
\end{equation*}

According to these propositions, it is enough to check the value $x_{m}$ $%
(M_{m})$ around the above $m_{0}$ when $q$ is $\frac{1}{2^{n}}$ for a large $%
n$. More generally, when $q$=$\frac{k}{2^{n}}$ with some integer $k,$ it is
easily checked that the above two propositions are held and the value $x_{m}$
$(M_{m})$ becomes over $\frac{1}{2}$ around the $m_{0}$ above.

One can think about various possible implementations of the idea of using
chaotic dynamics for computations, which is open and very intersting
problem. About this problem, realization of nonlinear quantum gates will be
essential, on which we will discuss in atomic quantum computer in \cite{OV2}.

Finally we show in Fig.1
how we can easily amplify the small $q$ in several
steps.

\section{Conclusion}

The complexity of the quantum algprithm for the SAT problem has been
considered in \cite{OM} where it was shown that one can build the unitary
matrix $U_{f}$ in the polynomial time. We have also to consider the number
of steps $m$ in the classical algorithm for the logistic map performed on
quantum computer. It is the probabilistic part of the construction and one
has to compute several times to be able to distingish the cases $q=0$ and $%
q>0.$ Thus it concludes that the quantum chaos algorithm can solve the SAT
problem in polynominal time according to the above propositions.

In conclusion, in this paper the quantum chaos algorithm is proposed. It
combines the ordinary quantum algorithm with quantum chaotic dynamics
amplifier. We argued that such a algorithm can be powerful enough to solve
the \textbf{NP}-complete problems in the polynomial time. Our proposal is to
show existence of algoritm to solve NP-complete problem. The physical
implimentation of this algorithm is another question and it will be strongly
desirable to be studied.

\newpage
%\FRAME{dtbpF}{8.9996cm}{6.0012cm}{0pt}{}{}{logistic.eps}{\special{language
%"Scientific Word";type "GRAPHIC";display "USEDEF";valid_file "F";width
%8.9996cm;height 6.0012cm;depth 0pt;original-width 52.75pt;original-height
%35.9375pt;cropleft "0";croptop "1";cropright "1";cropbottom "0";filename
%'logistic.eps';file-properties "NPEU";}}
\begin{figure}
\begin{center}
\includegraphics*[width=8cm]{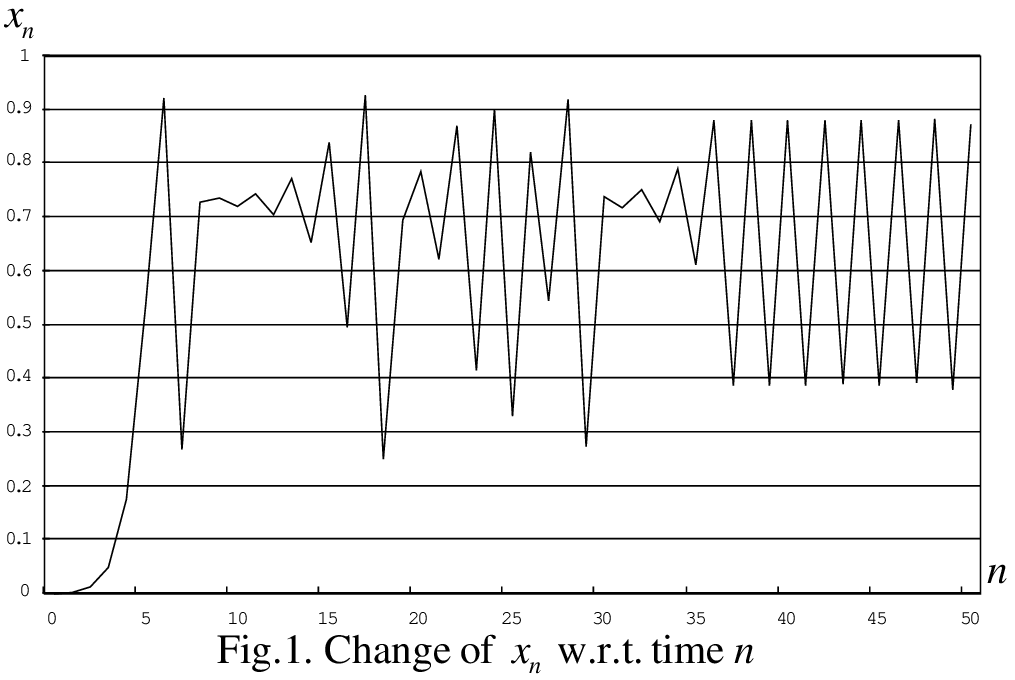}
%\epsfile{file=logistic.eps,width=8cm}
\end{center}
\label{fig:log}
\end{figure}
\end{document}